\documentclass[prd,aps,floats,twocolumn,nofootinbib]{revtex4-1}
\usepackage{slashed}
\usepackage{mathtools}
\usepackage{amsfonts}
\usepackage{amssymb}
\usepackage{epsfig}

\begin{document}

\newcommand{\m}[1]{\mathcal{#1}}
\newcommand{\nn}{\nonumber}
\newcommand{\ph}{\phantom}
\newcommand{\eps}{\epsilon}
\newcommand{\be}{\begin{equation}}
\newcommand{\ee}{\end{equation}}
\newcommand{\bea}{\begin{eqnarray}}
\newcommand{\eea}{\end{eqnarray}}
\newtheorem{conj}{Conjecture}
\newcommand{\plk}{\mathfrak{h}}

%%%%%%%%%%%%%%%%%%%%%%%%%%%%%

%\title{wave function of the Universe with a variable cosmological constant}
\title{Minisuperspace Quantum Cosmology from the Einstein-Cartan Path Integral}
\date{ }

\author{Raymond Isichei}
\author{Jo\~{a}o Magueijo}
\email{j.magueijo@imperial.ac.uk}
\affiliation{Theoretical Physics Group, The Blackett Laboratory, Imperial College, Prince Consort Rd., London, SW7 2BZ, United Kingdom}

\begin{abstract}
We derive the fixed-$\Lambda$ and unimodular propagators using the path integral formalism as applied to the Einstein-Cartan action. The simplicity of the action (which is linear in the lapse function) allows for an exact integration starting from the lapse function and the enforcement of the Hamiltonian constraint, leading to a product of Chern-Simons states if the connection is fixed at the endpoints. No saddle point approximation is needed. Should the metric be fixed at the endpoints, then, depending on the contour chosen for the connection, Hartle-Hawking or Vilenkin propagators are obtained. Thus, in this approach one trades a choice of contour in the lapse function for one in the connection, where appropriate. The unimodular propagators are also trivial to obtain via the path integral, and the previously derived expressions are recovered.  \end{abstract}

\maketitle

\section{Introduction}
Path integral quantisation is currently the choice method for obtaining a wavefunction of the Universe in quantum cosmology. While boundary conditions of the Universe are ultimately a matter of choice, the path integral is an ideal setting for the discussion and implementation of these boundary conditions as they naturally appear as limits of integration. The path integral of the Einstein-Hilbert action in the ADM formalism is very closely related to Dirac canonical quantisation in which constraints are operators which annihilate wavefunction eigenstates. In the ADM formalism, the Einstein-Hilbert action contains three momentum constraints and the Hamiltonian constraint. Hartle \& Hawking observed that by taking the path integral of the Einstein-Hilbert action containing Hamiltonian constraint, the wavefunction of the Universe $\Psi$ obeys the Wheeler-DeWitt equation~\cite{HH1,H1}
\begin{equation}
    \mathcal{H}\Psi=0.
\end{equation}
The three Hamiltonian constraints correspond to the wavefunction being invariant under spatial diffeomorphisms. If homogeneity and isotropy are assumed, then the shift function $N^{i}=0$ and the only relevant constraint is the Hamiltonian constraint. Halliwell rigorously showed that the minsuperspace FRW Einstein-Hilbert action can be made BRST invariant and gauge fixed such such that the Lapse function $N$ can be chosen to be constant~\cite{Hal1,Jonathan}.\\

The purpose of this paper is to explore the relationship between path integral and canonical quantisation. We want to show that recent results based on the canonical quantisation of the Einstein-Cartan action~\cite{JoaoLetter,JoaoPaper} can also be derived by taking the path integral of the Einstein-Cartan action for the FRW metric. In this formalism, the connection is more fundamental than the metric~\cite{kodama} and unimodular extensions~\cite{unimod,unimod1} take center place.

In such theories a dual relation can be found between the Hartle-Hawking wavefunction and the Chern-Simons state, should the connection be real. Allowing for a negative imaginary connection contour makes contact with the Vilenkin wavefunction instead~\cite{CSHHV}. The unimodular extension, rendering $\Lambda$ constant on-shell only, permits superposing such waves into localized packets. These are normalizable under an inner product suggested by the unimodular theory, with unitarity preserved under evolution as ticked by a clock associated with the variable conjugate to $\Lambda$.

The propagators for such theories have been worked out in~\cite{HHpackets}. Whether the concept of propagator is useful in discussing creation out of nothing in the context of this quantum theory is a different matter. Regardless of that issue, the path integral formulation should help to sharpen up the derivation of these propagators and clarify the discussion.

\section{Previous results}
The starting point here is the 
Einstein-Cartan action reduced to minisuperspace (MSS):
\begin{equation}\label{Sg}
S_0=\frac{3V_c}{8\pi G} \int 
dt\bigg(\dot{b}a^2-Na \left[- (b^{2}+k 
%-c^2 -\frac{\Lambda }{3}%
%a^{2}\bigg)
)+ \frac{a^2}{\phi}\right]\bigg).
%dt\bigg(a^{2}\dot{b}+Na (b^{2}+k +...
%-c^2 -\frac{\Lambda }{3}%
%a^{2}\bigg)
%)\bigg).
\end{equation}
where $a$ is the expansion factor, $b$ is the only MSS connection variable (an off-shell version of the Hubble parameter, since $b= \dot a$ on-shell, if there is no torsion), $k$ is the normalized spatial curvature, $N$ is the lapse function and $V_c=\int d^3 x$ is the comoving volume of the region under study ($V_c=2\pi^2$ for a whole $k=1$ Universe) and $\phi=3/\Lambda$ is a useful parameterization for the cosmological constant. Note that in this formulation the natural canonical pair consists of $b$ and $a^2$, the MSS reduction of the connection (gauge field) and the densitized inverse triad (the electric field), respectively.  

We then subject this theory to the unimodular extension~\cite{unimod1} in the Henneaux and Teitelboim formulation~\cite{unimod}, where full diffeomorphism invariance is preserved. To $S_0$ one adds a new term:
\be\label{Utrick}
S_0\rightarrow S=
S_0- \frac{3}{8\pi G }  \int d^4 x \, \phi\,  \partial_\mu T^\mu
%=S_0+ \int d^4 x (\partial_\mu\Lambda)  T^\mu_U
\ee
where $T^\mu$ is a  density, so that the added term is diffeomorphism invariant without the need for a $\sqrt{-g}$ factor in the volume element or for the connection in the covariant derivative. Since the metric and connection do not appear in the new term, the Einstein equations and other 
field equations are unmodified. 
The only new equations of motion are
the  on-shell constancy for $\Lambda$ (the defining characteristic of unimodular theories~\cite{unimod1,unimod}) and the fact that $T^0$ is proportional to a prime candidate for relational time: 4-volume or unimodular time~\cite{unimod,unimod1,Bombelli,UnimodLee2}.
We stress that the pre-factor in the new term is arbitrary and chosen for later convenience, but it does matter for the quantum theory. Likewise we could have defined the alternative action:
\be\label{Utrickalt}
S_0\rightarrow S=
S_0+ \frac{3}{8\pi G }  \int d^4 x \, (\partial_\mu\phi)\,   T^\mu
%=S_0+ \int d^4 x (\partial_\mu\Lambda)  T^\mu_U
\ee
equivalent to (\ref{Utrick}) up to boundary term:
\begin{equation}
    S_{\rm bound}=-\frac{3}{8\pi G }  \int_{\partial {\cal M}}d\Sigma_\mu \, \phi\,   T^\mu,
\end{equation}
where $d\Sigma_\mu$ is the coordinate area density normal on the boundary $\partial {\cal M}$. 
This does not matter classically and for the purpose of canonical quantization, but it does affect the path integral formulation, as we shall see. As it happens, (\ref{Utrick}) is the correct form of the action if we wish to use $T$ to label the starting and ending times for the propagators. 
Reduction to MSS gives:
\be\label{Utrickmss}
S_0\rightarrow S=
S_0 +  \frac{3 V_c}{8\pi G }  \int dt x \, \dot \phi\,  T
%=S_0+ \int d^4 x (\partial_\mu\Lambda)  T^\mu_U
\ee
(with $T\equiv T^0$). Note that on-shell we have:
\begin{equation}
\dot T =N\frac{ a^3}{\phi^2}=N\frac{\Lambda^2}{9}a^3,
\end{equation}
proportional to 4-volume or unimodular time~\cite{Bombelli,UnimodLee2}.

The quantum mechanics of this theory was studied in~\cite{JoaoPaper} in the connection representation and in~\cite{HHpackets} in the metric representation. In the latter the propagators were worked out directly. The general solutions are superpositions of fixed-$\Lambda$ solutions to the WDW equation, $\psi_s$, augmented by a time evolution factor:
\bea
\psi(q,T)&=&
\int^\infty_{-\infty} %\frac{d\phi}{\sqrt{2\pi\plk}}
d\phi\, 
{\cal A}(\phi)
e^{-\frac{i}{\plk}\phi T} \psi_s(q;\phi)
\label{HHPacketsq}
\eea
where  $q$ stands for one of the dual variables, $b$ or $a^2$, and $\plk=l_P^2/(3V_c)$ with $l_P=\sqrt{8\pi G\hbar}$. For $q=b$:
\begin{equation}
    \psi_s(b,\phi)=\psi_{CS}(b,\phi)={\cal N}_b\exp{\left[\frac{i}{\plk } \phi X (b) \right]},
\end{equation}
i.e. the Chern-Simons-Kodama state~\cite{jackiw,kodama,witten,lee1,lee2,Randono0,Randono1,Randono2,Wieland,RealCS}, where 
\begin{equation}
    X(b)=\frac{b^3}{3}+kb
\end{equation}
is the Chern-Simons functional reduced to MSS. 
For $q=a^2$ one has for $\Lambda>0$:
\begin{eqnarray}
\psi_s\left(a^{2},\phi\right)&=&
{\cal N}_a {\rm Ai} (-z)\label{psisAi}
\end{eqnarray}
with 
\begin{eqnarray}
z&=&-\left(\frac{\phi}{\plk}\right)^{2/3}
\left(k-\frac{ a^2}{\phi}\right), 
\end{eqnarray}
i.e. the Hartle-Hawking wavefunction, or for $\Lambda<0$:  
\begin{eqnarray}\label{psisaibi}
\psi_s\left(a^{2},\phi \right)&=&
\frac{{\cal N}_a}{2} [{\rm Ai} (-z)+ i{\rm Bi} (-z)]
\end{eqnarray}
i.e. the Vilenkin counterpart. A central assumption here is that $b$ is real, so that these $\psi_s(a^2;\phi)$ are the Fourier duals of the CSK state~\cite{CSHHV,HHpackets}. (The normalization factors ${\cal N}$ defined in~\cite{HHpackets} will not be important here. The apparent contradiction with~\cite{JLnegL}  for $\Lambda<0$ is currently being investigated.) 

If instead we allow for $b$ to have a contour covering half the negative imaginary line and the positive imaginary line, then~\cite{CSHHV} we find the V wavefunction for $\Lambda>0$, and the HH wavefunction for $\Lambda<0$. The unimodular wave packets for such a theory, however, have not been studied, since the Fourier duality must then be replaced by a Laplace transform, with technical complications~\cite{HHpackets}.

The reason why unimodular propagators can be found directly is that the  (time-independent) amplitude ${\cal A}(\phi)$ can be obtained by an inversion formula:
\begin{eqnarray}\label{ampa2b}
    {\cal A}(\phi) &=&\int  d\mu(q)\, e^{\frac{i}{\plk}\phi T} \psi^\star_s  (q;\phi) \psi (q,T),
\end{eqnarray}
with $d\mu(b)=dX(b)$ and $d\mu(a^2)=da^2\, a^2/\phi$. Hence we can write (\ref{HHPacketsq}) in the form:
\begin{equation}\label{Greendef}
    \psi(q,T)=\int d\mu(q') 
    %G(q,T;q',T')
    \langle q,T |q',T'\rangle
    \psi(q',T')
\end{equation}
reading off the propagator as:
\begin{eqnarray}\label{unimodprop}
\langle q,T |q',T'\rangle&=&
   \int d\phi\, e^{-\frac{i}{\plk } \phi \Delta T} \langle q|q' \rangle_\phi 
\end{eqnarray}
related to the fixed-$\Lambda$ propagator:
\begin{equation}\label{propL}
    \langle q|q' \rangle_\phi=
    %\langle q|\phi\rangle\langle \phi |q' \rangle= 
    \psi_s(q;\phi) \psi_s^\star (q';\phi). 
\end{equation}
This is the central result we will attempt to reproduce using the path integral formalism. We can have $b$ or $a^2$ at both endpoints, or mixed propagators. 

We start with the fixed-$\Lambda$ propagators with fixed connections as endpoints.

\section{The connection fixed-$\Lambda$ propagators}
We want to compute:
\begin{equation}
    Z_0=\int e^{\frac{i}{\hbar}S_0}
\end{equation}
with fixed $b=b_f$ and $b=b_i$ as endpoints. This is such a simple system that the answer can be obtained in a variety of ways. For the sake of diversity we present an approach that has not been used in previous literature. The idea is to integrate the unconstrainted lapse function first, leading to a delta Dirac in the Hamiltonian constraint. In contrast with~\cite{Jonathan} this leads to: 1) using solutions to the Hamiltonian constraint for the rest of the integration, 2) eschewing the need for a saddle point approximation. In analogy with~\cite{JeanLuc1} we will be using Lorentzian path integrals, but will not need the sophisticated machinery developed therein. In contrast with all previous approaches we will assume that $N$ is real and covers the whole real line. Freedom in the choice of contours will come later, as we will see in Section~\ref{mixedprops}.

An elegant derivation of the connection propagators follows, with one caveat. 
Setting $\tilde N=Na$ and $H=-(b^2+k)+a^2/\phi$, we get: 
\begin{eqnarray}
\label{CS}
    \langle b_f|b_i\rangle_\phi  &=& \int  {\cal D} b \, {\cal D} a^2 \, {\cal D}\tilde N \, \exp{\left[\frac{i}{\plk}\int dt\, \left(\dot b a^2 -\tilde  N H \right)\right]}\nn\\
    &=&\int  {\cal D} b \, {\cal D} a^2 \,  \exp{\left[\frac{i}{\plk}\int dt\, \dot b a^2  \right]}\delta[H]\nn\\
    &\propto &\int  {\cal D} b \,   \exp{\left[\frac{i}{\plk}\int dt\, \phi \, \dot b (b^2+k)\right]}\nn\\
    &= &\int  {\cal D} b \,   \exp{\left[\frac{i}{\plk}\int^{b_f}_{b_i} db\, \phi \, (b^2+k)\right]}\nn\\
    &\propto& \psi_{CS}(b_f,\phi)\psi^\star _{CS}(b_i,\phi),
\end{eqnarray}
where in the last step we used:
\begin{equation}
    \int^{t_f}_{t_i} dt\, \dot b (b^2+k)=\int^{b_f}_{b_i} db (b^2+k)=X(b_f)-X(b_i)\nn 
\end{equation}
and the functional integration over $a^2$ in the second step includes integrations over $a^2_f$ and $a^2_i$, so that the Hamiltonian constraint is also imposed there.

The caveat is that the proportionality constant in the last step is infinite. This is not surprising. 
%since we kept an infinite gauge volume inside the integral. 
Given the invariance of the theory under redefinitions of time/lapse function, the  path integral has an infinite gauge volume multiplying our answer. But this is not a problem and can be eliminated in at least two ways.

\subsection{Delayed gauge fixing}
One possibility is to fully fix the gauge. But if we want to implement the Hamiltonian constraint as in the calculation leading to (\ref{CS}), we should write the gauge-fixing condition as a functional integral over an auxiliary field $B$:
\begin{equation}
    \delta[\tilde N-\tilde N_0]=\int {\cal D} B \; e^{\frac{i}{\plk}\int dt\, B(\tilde N-\tilde N_0)},
\end{equation}
and then ``delay'' this gauge fixing by performing the integration in $B$ last. (Here we did the gauge fixing with $\tilde N=a N$ but this would have worked with any other version of $N$). Specifically:
\begin{eqnarray}
    \langle b_f|b_i\rangle_\phi  &=& \int {\cal D} B\, {\cal D} b \, {\cal D} a^2 \, {\cal D}\tilde N \, \nn\\
    &&\exp{\left[\frac{i}{\plk}\int dt\, \left(\dot b a^2 -\tilde  N (H-B) \right)-B\tilde N_0\right]}.\nn
\end{eqnarray}
The integration over $\tilde N$ now produces the functional Dirac delta:
\begin{equation}
   \delta[H-B]= \delta\left[ -(b^2+k) +\frac{a^2}{\phi} -B\right]\, 
\end{equation}
where before we had simply $\delta[H]$. This is 
eaten up by the functional integration in $a^2$ to produce:
\begin{eqnarray}
    \langle b|b'\rangle_\phi &=& \int {\cal D} B\, {\cal D} b \, \label{intermediate-result}\\
    &&\exp{\left[\frac{i}{\plk}\int dt\, \left(\dot b
    \phi(b^2+k)+\dot b \phi B-B\tilde N_0\right)\right]}.\nn
  \end{eqnarray}  
The first term in the exponent in (\ref{intermediate-result}) produces boundary factors that are the right result, Eq.~(\ref{propL}), just as in the calculation (\ref{CS}). However, the proportionality factor in the last step in (\ref{CS}) is now:
\begin{eqnarray}
     &&\int {\cal D} B\, {\cal D} b \, \exp{\left[\frac{i}{\plk}\int dt\, B\left(\dot b \phi -\tilde N_0\right)\right]}\nn\\
     &=& \int {\cal D} b \, \delta(\dot b \phi-\tilde N_0)\nn\\
     &=&1\nn
\end{eqnarray}
with  $\dot b=\tilde N_0/\phi={\rm const}$ and so:
\begin{equation}
    \frac{b_f-b_i}{t_f-t_i}=\frac{\tilde N_0}{\phi}
\end{equation}
enforced. 
%(Throughout this calculation we have ignored irrelevant numerical/constant factors.) Notice that 
This is just a statement of the Raychaudhuri equation in this gauge:
\begin{equation}
    \dot b=\frac{\Lambda}{3}a N
    =\frac{{\tilde N}_0}{\phi}.
\end{equation}
Hence we have 
\begin{equation}
 \langle b_f|b_i\rangle_\phi \propto \psi_{CS}(b_f,\phi)\psi^\star _{CS}(b_i,\phi),
\end{equation}
as required, with the proportionality sign now hiding no infinities. 

\subsection{Ratios of Path Integrals}
The infinite prefactor multiplying the Chern-Simons functional in (\ref{CS}) can also be dealt with by taking the ratio of the path integral containing the Hamiltonian and the the path integral for the classical action given by
\begin{equation}
    Z_{cl}=\int \mathcal{D}\Tilde{N} \mathcal{D}b\mathcal{D}a^{2} \ \text{{exp}} \Biggl[\frac{i}{\plk}\int dt \ a^{2}\dot{b} \Biggl].
\end{equation}
Using the fact that the classical equation of motion for $b$ is the Raychaudhuri equation, this can be substituted into the classical action to yield
\begin{align}
    \begin{split}
        Z_{cl}&=\int \mathcal{D}\Tilde{N} \mathcal{D}b\mathcal{D}a^{2} \ \text{{exp}} \Biggl[\frac{i}{\plk}\int dt \ a^{2}\frac{\Lambda}{3}aN \Biggl],\\
        &=\int \mathcal{D}\Tilde{N} \mathcal{D}b\mathcal{D}a^{2} \ \text{{exp}} \Biggl[\frac{i}{\plk}\int dt \ a^{2}\frac{\Lambda}{3}\Tilde{N} \Biggl],\\
        &=\int  \mathcal{D}b\mathcal{D}a^{2} \ \delta \biggl[\biggl(\frac{\Lambda}{3}a^{2}\biggl)\biggl],\\
        &=\int \mathcal{D}b.
    \end{split}
\end{align}
%It is important to note that the substitution of the Raychaudhuri equation is not without motivation. As shown previously, if the path integral for $a^{2}$ is evaluated first, this gives a delta function containing the classical equations of motion. The original path integral is only non vanishing when the Raychaudhuri equation holds. 
Taking the ratio of the path integrals then gives us a product of Chern-Simons functions for these boundary conditions: 
\begin{align}
\label{ratio}
\begin{split}
\frac{Z_{0}}{Z_{cl}}&= \frac{\int \mathcal{D}\Tilde{N}\mathcal{D} b\mathcal{D}a^{2} \ \text{{exp}}\biggl[\frac{i}{\plk} \int dt \ \biggl(\dot{b}a^{2}-\Tilde{N}H \biggl) \biggl] }{\int \mathcal{D}\Tilde{N}\mathcal{D} b\mathcal{D}a^{2} \ \text{{exp}}\biggl[\frac{i}{\plk}\int dt \ \biggl(\dot{b}a^{2} \biggl) \biggl] },\\
&=\frac{\int \mathcal{D}b}{\int \mathcal{D}b} \ \psi_{CS}(b_{f},\phi)\psi^{*}_{CS}(b_{i},\phi),\\
&=\psi_{CS}(b_{f},\phi)\psi^{*}_{CS}(b_{i},\phi).
\end{split}
\end{align}
The Chern-Simons function product is a constant with respect to the intermediate $b$ and is therefore independent of the path integral on the second line of (\ref{ratio}). The divergence is then cancelled by the fact that the path integral for the classical action yields the same integral.
\section{The other fixed-$\Lambda$ propagators}\label{mixedprops}
The calculation is subtly different for propagators involving the metric at either or both endpoints, since then the action (\ref{Sg}) no longer reproduces the equations of motion.  
Recall that for any theory ruled by a phase space action principle~\cite{Goldstein}:
\begin{equation}
    S=\int dt \ (\dot q p -H(q,p))
\end{equation}
the action only provides the equations of motion if the variable which is dotted in the first term (the ``$q$'') is kept fixed at the endpoints. Otherwise, boundary counter-terms have to be added as appropriate. For example, if $q$ is allowed to vary at both endpoints we should use the alternative action: 
\begin{equation}
    S\rightarrow S-[qp]^f_i. 
\end{equation}
In contrast, $p$ does not need to be kept fixed. No boundary terms are required, should it be allowed to vary at the endpoints. This is completely general, and has counterparts in action principle formulations of gravity.

This point was noted in~\cite{Jonathan} in the context of the Einstein-Hilbert action, where the boundary counter-term is a MSS version of the Gibbons-Hawking-York boundary term~\cite{Boundary1,Boundary2} (see also~\cite{JeanLucBoundary} for a more general discussion). If our starting point is the Einstein-Cartan action (\ref{Sg}), then $b$ is the variable to be kept fixed (the one dotted in the first term of (\ref{Sg})), unless boundary terms are added. 
But propagators are predicated on the fact that if a variable is fixed at one endpoint, then its dual is implicitly fully unconstrained there. In order to obtain the propagators fixing the metric at an endpoint, we therefore need to add a corresponding counter-term to (\ref{Sg}), since at that endpoint $b$ must necessarily be left unfixed.

In the Einstein-Cartan formulation, we only need an integration by parts when varying with respect to $b$ to obtain the torsion-free condition:
\begin{equation}
\frac{\delta S_0}{\delta b} = 0 \implies    \dot a=Nb.
\end{equation}
Such a variation generates a boundary term of the form:
\begin{equation}
    \delta S_0=({\rm volume\; term}) +\frac{3V_c}{8\pi G}[a^2 \delta b]^f_i
\end{equation}
A boundary counter-term proportional to $a^2 b$ should therefore be added for each endpoint (start, finish or both) where $a^2$ is fixed and $b$ is left unconstrained. Once again, this counter-term is a MSS version of the general counter-term for Einstein-Cartan theory, identical on-shell to the Gibbons-Hawking-York boundary term (see e.g.~\cite{Boundary3,Boundary4}). The crucial point here is that in the Einstein-Cartan formalism\footnote{And its offshoots, such as the Ashtekar or Plebanski formalisms.} it is the connection that must be kept fixed if we want to dispense with boundary counter-terms.

For example, for computing $\langle a^2|b'\rangle$ we should replace (\ref{Sg}) by: 
\begin{equation}\label{newSg}
    S_0\rightarrow S_0-\frac{3V_c}{8\pi G} b_f a_f^2. 
\end{equation}
The calculation in the last Section still carries through but at the end we are left with a simple (non-functional) integral over $b_f$ of the form:
\begin{eqnarray}
 \langle a^2_f|b_i\rangle_\phi &\propto &\int db_f\,  \exp{\left[-\frac{ib_fa^2_f}{\plk}\right]} \psi_{CS}(b_f,\phi)\psi^\star _{CS}(b_i,\phi),\nn
\end{eqnarray}
where the exponential reflects the new term in (\ref{newSg}). Hence we end up with the integral transform of the Airy functions, the contour in $b_f$ now deciding which one we get. Assuming  $\Lambda>0$, and a real $b$ contour~\cite{CSHHV}, we have:
\begin{equation}
    \langle a^2_f|b_i\rangle_\phi \propto {\rm Ai}(-z_f)\psi^\star _{CS}(b_i,\phi),
\end{equation}
that is, the correct version of (\ref{propL}), with the other variations obtained by a different choice of contour and sign of $\Lambda $. This propagator between fixed initial $b$ and final fixed $a^{2}$ is identical to the propagator previously derived in~\cite{Jonathan}.

Likewise for all the other cases of propagators involving $a^2$. As a final example, consider $\langle a^2_f|a^2_i\rangle_\phi $. The action should be 
\begin{equation}
    S_0\rightarrow S_0-\frac{3V_c}{8\pi G} [b a^2]^f_i. 
\end{equation}
so we end up with 
\begin{eqnarray}
 \langle a^2_f|a^2_i\rangle_\phi &\propto &\int db_f\,  \exp{\left[-\frac{ib_fa^2_f}{\plk}\right]} \psi_{CS}(b_f,\phi) ,\nn\\
 &&\times \int db_i\,  \exp{\left[\frac{ib_ia^2_i}{\plk}\right]} \psi^\star _{CS}(b_i,\phi),\nn
\end{eqnarray}
which for $\phi>0$ and real $b$ leads to
\begin{equation}
     \langle a^2_f|a^2_i\rangle_\phi 
      \propto {\rm Ai}(-z_f){\rm Ai}(-z_i). 
\end{equation}

As explained in~\cite{HHpackets}, if $\Lambda<0$, then for a real $b$ we must replace the HH wavefunctions by their V counterpart. The situation is reversed if we give $b$ the contour $(-i\infty, 0) \ \bigcup \ (0, \infty)$ in the complex $b$ plan, that is, the negative imaginary line and the positive real line. Then, we get the V wavefunction for $\Lambda>0$ and the HH one for $\Lambda>0$, and this change must be made in all mixed propagators. For example, we get
\begin{align}
    \begin{split}
       \langle a^2_f|a^2_i \rangle_\phi & \propto [\text{{Ai}}(-z_{f})+i\text{{Bi}}(-z_{f})]\\  
       & \times [\text{{Ai}}(-z_{i})+i\text{{Bi}}(-z_{i})]
    \end{split}
\end{align}
for the metric propagator if $\Lambda<0$.

Thus, we have replaced a choice of integration contour in $N$ (as in~\cite{Jonathan,JeanLuc1,JeanLuc2}) by a choice of contour in the connection $b$. 

%[SHALL WE ADD THE VILENKIN CASE HERE? IT DOES NOT SEEM TO MATTER AT THIS POINT, BUT IT MIGHT LATER, FOR THE UNIMODULAR PACKETS]

\section{Implication for creation from nothing}

The fact that our calculations did not use the saddle point approximation has a simple implication for ``creation from nothing''. 
Setting $z_0\equiv z(a^2=0)=-(\phi/\plk)^{2/3}k$, this is usually expressed by the amplitude:
\begin{eqnarray}\label{ampvacstar}
 \langle  a^2=a^2_\star|a^2=0\rangle_\phi&\propto & {\rm Ai}(0) {\rm Ai}(-z_0)\nn\\
 &\propto &\exp{\left[-\frac{2}{3} \frac{\phi}{\plk}k^{3/2}\right]}\nn\\
& =& \exp{\left[-\frac{6 V_c k^{3/2}}{l_P^2\Lambda}\right]}
\end{eqnarray}
for the HH case, and
\begin{eqnarray}\label{ampvacstar1}
 \langle  a^2=a^2_\star|a^2=0\rangle_\phi&\propto & 
 %({\rm Ai}(0)+i\text{{Bi}}(0)) (
 {\rm Ai}(-z_0)+i\text{{Bi}}(-z_0)\nn\\
 &\propto &\exp{\left[\frac{2}{3} \frac{\phi}{\plk}k^{3/2}\right]}\nn\\
& =& \exp{\left[\frac{6 V_c k^{3/2}}{l_P^2\Lambda}\right]}
\end{eqnarray}
for the V case, where the WKB approximation was used in the second line of both calculations. We will not comment on the implications of the exponent's sign for inflationary models (which are not the topic of this paper).  

We want to note here that we could equally well have identified the ``nothing'' from:
\begin{equation}
    - (b^{2}+k 
%-c^2 -\frac{\Lambda }{3}%
%a^{2}\bigg)
)+ \frac{a^2}{\phi}=0 \implies (a=0\iff b=\pm i \sqrt{k})
\end{equation}
resulting in $b=\pm i \sqrt{k}$. Neither of these
$b$ points lie on the HH contour but  $b=- i \sqrt{k}$ is on the contour leading to the V wave function. Hence in this case we could also have represented the ``creation from nothing'' by the propagators:
\begin{eqnarray}\label{ampvacstar2}
 \langle  a^2=a^2_\star|b=-i\sqrt{k}
 \rangle_\phi&\propto & \psi_{CS}^\star (b=-i\sqrt{k})\nn\\
 &\propto & \exp{\left[\frac{6 V_c k^{3/2}}{l_P^2\Lambda}\right]}
\end{eqnarray}
or:
\begin{eqnarray}\label{ampvacstar3}
 \langle  b=0 |b=-i\sqrt{k}
 \rangle_\phi&\propto & \psi_{CS}^\star (b=-i\sqrt{k})\nn\\
 &\propto & \exp{\left[\frac{6 V_c k^{3/2}}{l_P^2\Lambda}\right]}.
\end{eqnarray}
Choosing $b=0$ or $a=a_\star$ as the nearest point of the classical trajectory makes little difference. However, choosing $a=0$ and $b=-i\sqrt{k}$ cannot be exactly equivalent. Indeed it only makes sense for the Vilenkin contour. Also we have not used the saddle point approximation in our propagator derivation; yet it is only after using the WKB approximation in (\ref{ampvacstar1})  that the it becomes equivalent to (\ref{ampvacstar2}).

The use of $b=-i\sqrt{k}$ as representative of the ``nothing'' has been advocated at least since~\cite{Louko}.

\section{The unimodular path integral and its propagators}

Given the simple form of the unimodular extension (\ref{Utrickmss}), its propagators (\ref{unimodprop})
can be recovered from the path integral by writing:
\begin{equation}\label{Zunimod}
    Z=\int {\cal D}\phi\, {\cal D}T \, \exp{\left[\frac{i}{\plk}\int dt\, \dot \phi
T    \right]} Z_0.
\end{equation}
Ambiguities in the classical theory, which are not relevant in the canonical quantization, stand out here: foremost the issue of the boundary terms and which of the two formulations, Eqs.~(\ref{Utrick}) or (\ref{Utrickalt}), to take.

It turns out that (\ref{Utrickalt}) is the correct choice, if we want to use relational time $T$ to index the start and end points of the propagators, and so leave $\phi$ totally unconstrained. 
If we take (\ref{Utrickalt}), then $\phi$ is the dotted variable, so by fixing relational time at (both) endpoints, we have add to the action a boundary term, with the full unimodular action reading:
\be\label{UtrickmssBT}
S=
S_0 +  \frac{3 V_c}{8\pi G }  \int dt x \, \dot \phi\,  T - \frac{3 V_c}{8\pi G }  [\phi T]^f_i
%=S_0+ \int d^4 x (\partial_\mu\Lambda)  T^\mu_U
\ee
and this will generate the necessary boundary terms in (\ref{Zunimod}). 

The integral is easy to do. Functional integration over $T$ leads to $\delta[\dot \phi]$. The functional integral in $\phi$ therefore requires that $\phi$ be constant along the intermediate trajectories, its (constant) value left undefined. The final result is therefore a (non-functional) integration over the same $\phi$ at both endpoints, which, taking into account the boundary terms in (\ref{UtrickmssBT}) is:
\begin{equation}
    Z=\int d\phi\, \exp{\left[-\frac{i}{\plk}\phi (T_f-T_i)\right]} Z_0
\end{equation}
thereby reproducing (\ref{unimodprop}). 

As in~\cite{HHpackets}, more explicit forms of the propagators can then be found. 
Assuming that the the $a^{2}$ and $N$ path integrals have been evaluated in $Z_0$ to yield the product of initial and final Chern-Simons functions, the unimodular propagator becomes:
\begin{eqnarray}
      Z&=&\int d\phi \ \exp{\left[-\frac{i}{\plk}\phi (T_f- X_f-T_i + X_i)\right]}\nn \\
        &=& \delta(X_f-T_f-X_i+T_i)=
        \delta(X^f_{\rm ret} - X^i_{\rm ret} )
\end{eqnarray}
with $X_{{\rm ret}}(b, T)\equiv X(b)- T$. This defines $b_{\rm ret}=b(X_{\rm ret})$ in terms of which we can write mixed propagators (associated with the other $Z_0$ as computed in Section~\ref{mixedprops}) in the alternative forms:
\begin{eqnarray}
\langle b T|a^2  T' \rangle&=&\frac{e^{\frac{i}{\plk }b_{\rm ret} a^2}} {\sqrt{2\pi \plk}(b_{\rm ret}^2+k)},\\
%\\ \langle a^2 T| b  T' \rangle&=&\frac{e^{-\frac{i}{\plk }b_{\rm adv} a^2}} {\sqrt{2\pi \plk}(b_{\rm adv}^2+k)},
 \langle a^2 T|a^{2\prime} T' \rangle
 &=&\int \frac{db}{2\pi\plk}\frac{e^{-\frac{i}{\plk}(b a^2-b_{ret}a^{2\prime})}}{b_{\rm ret}^2+k}
 %\nn\\ &=&\int  \frac{db_{\rm ret}}{2\pi\plk} \frac{e^{-\frac{i}{\plk}(b a^2-b_{ret}a^{2\prime})}}{b^2+k}
\end{eqnarray}
with the latter simplifying to:
\begin{equation}\label{eqtpropa2}
    \langle a^2 0|a^{2\prime} 0 \rangle
    =\frac{e^{\frac{-|\Delta a^2|\sqrt{k}}{\plk}}}{2\sqrt{k}\plk}
\end{equation}
in the limit of equal times.

\section{Conclusions}
The core result of this paper is that one can eliminate the metric (or densitized inverse triad, $a^2$) in the path integral by integrating first over the unconstrained lapse function to impose the Hamiltonian constraint, which can then be used to write the metric as a function of the connection. The Einstein-Cartan action thus becomes a pure boundary term in the connection, identical with the Chern-Simons functional, so that the path integral with fixed connection at both endpoints is just the product of Chern-Simons-Kodama wavefunctions at endpoints and a redundant infinite integration over the intermediate connections (which we regularized)\footnote{Related results can be obtained from a path integral definition of ``pure connection'' formulations of General Relativity in which the metric is absent altogether; see~\cite{steffen}.}.

If one or both endpoints have fixed metric instead of connection, then we are left with an integration over the connection at that point and a path integral factor (arising from a boundary term in the action), which amounts to a Fourier/Laplace transform of the Chern-Simons-Kodama state. This converts the corresponding Chern-Simons-Kodama factor in the propagator into an appropriate Airy function, depending on the contour of the connection. The procedure thus yields Hartle-Hawking or Vilenkin propagators depending on the choice of contour for the connection (which has replaced the choice of contour for the lapse function). Therefore finding the propagators in connection space is much more practical, since the saddle-point approximation and steepest descent methods are required to find metric representation propagators. These calculations can be trivially extended to unimodular theory. 

Naturally, part of the reason for our successes is that the fixed-$\Lambda$ theory is very simple: it has zero degrees of freedom. But this is not true for the unimodular extension, which in minisuperspace has one degree of freedom (two more variables, the same number of constraints). 
%even though $\Lambda$ is a constant on-shell.
Unimodular theory is remarkable in that this happens without changing the number of {\it local} degrees of freedom. This is because locally we have a gauge symmetry (see~\cite{unimod}) and so an extra first-class constraint, so that the two new variables do not represent a new local degree of freedom. The homogeneous ``zero-mode'' of $\Lambda$ and of unimodular time are physical, but locally we still only have the two degrees of freedom of the standard graviton. Some of the techniques in this paper can also be applied to more complicated systems, including cosmologies with dust and radiation fluids subject to the ``unimodular trick''~\cite{JoaoLetter,JoaoPaper}. This is described in~\cite{raythesis}.

An important direction for future work following from our paper is the study of its implications for the stability of tensor mode perturbations around the homogeneous and isotropic background (as opposed to the scalar modes investigated in~\cite{Vilenkin-yamada}). We have in mind in particular the results of~\cite{JeanLuc1,JeanLuc2}. Given that these results were derived  within the path integral metric formalism, it would be interesting to investigate the issue from the point of view of the connection duals investigated here. One must wonder if the instability identified in~\cite{JeanLuc1,JeanLuc2} is 
related to similar issues plaguing some, but not all versions of the Chern-Simons-Kodama state~\cite{witten,lee2,dion,laura,RealCS,Steph-Laurent}. In general one finds unphysical states, with negative norms and/or energies, if the reality conditions are not properly taken into account. The Vilenkin state requires the connection to stray off the real line~\cite{CSHHV}, so the two instabilities may well be connected. It is also inevitable to wonder whether the picture on instabilities would be radically different within the unimodular extension. One must recall that, strictly speaking, no monochormatic wavefunction is physical and that this affects discussions of normalizability.

In closing, we stress that in this paper we have remained agnostic regarding the probability interpretation of the wave function. That is, we have used the propagators as a tool for evaluating the wave function, decoupling the matter from their probability interpretation. This is the attitude in~\cite{HHpackets}, Sec.~VII, as well as, one imagines, in~\cite{Vilenkin-yamada} (where presumably a Klein-Gordon current is to be used to interpret the wave function, rather than taking the square, as naively implied by the propagators). Obviously, the probability interpretation is very important, and this is addressed  elsewhere (see Ref.\cite{JoaoPaper}, Section VD, and Ref.~\cite{HHpackets}, sections III and V). The latter are applicable to the Hartle-Hawking wave function, but not to the Vilenkin tunneling wave function (a matter currently under investigation);  see also~\cite{Vilenkin-BC} for an important alternative view of unitarity and the correspondence principle. Whichever probability interpretation one takes, the technical aspects of our paper stand valid.

\section{Acknowledgments}
We thank Bruno Alexandre, Steffen Gielen, Jonathan Halliwell and Jean-Luc Lehners for discussions related to this paper. This work was supported by the STFC Consolidated Grant ST/T000791/1 (J.M.).

\end{document}